\documentclass[twocolumn,amsmath,amssymb,aps,pra]{revtex4}
\usepackage{bm}
\usepackage{amsfonts}
\usepackage{amsmath}
\usepackage{graphicx}
\usepackage{float}

\begin{document} 
\title{Knotted trajectories of neutral and charged particles in Gaussian light beams}
\author{Tomasz Rado\.zycki}
\email{t.radozycki@uksw.edu.pl}
\affiliation{Faculty of Mathematics and Natural Sciences, College of Sciences, Institute of Physical Sciences, Cardinal Stefan Wyszy\'nski University, W\'oycickiego 1/3, 01-938 Warsaw, Poland} 
\begin{abstract}
Making use of the equivalence between paraxial wave equation and two-dimensional Schr\"odinger equation, Gaussian beams of monochromatic light, possessing knotted nodal structures are obtained in an analytical way. These beams belong to the wide class of paraxial beams called the Hypergeometric-Gaussian beams [E. Karimi, G. Zito, B. Piccirillo, L. Marrucci and E. Santamato, Opt. Lett. {\bf 32}, 3053(2007)]. Four topologies are dealt with: the unknot, the Hopf link, the Borromean rings and the trefoil. It is shown in the numerical way that neutral polarizable particles placed in such light fields, upon precise tuning of the initial conditions, can be forced to follow the identical knotted trajectories. A similar outcome is also valid for charged particles that are subject to a ponderomotive potential. This effect can serve to precisely steer particles along chosen complicated pathways exhibiting non-trivial topological character, guide them around obstacles and seems to be helpful in engineering more complex nanoparticles. 
\end{abstract}
\maketitle

\section{Introduction}\label{int}

In recent years, it has proven possible to investigate and generate beams of light with some complex structure far from the academic concept of plain waves. Theoretical and experimental studies of ``structured'' light, ``non-diffracting'' or ``accelerating'' beams have been developed~\cite{arlt1,flo,sgg,siv,andrews,lee,zhang,babik} opening a variety of possible applications, in particular for trapping and guiding particles, atoms, molecules or even micrometer-sized objects. Among beams that have gained special interest one can enumerate Laguerre-Gaussian~\cite{lg,lg2,arlt2}, Bessel~\cite{arlt2,durnin1,durnin2,vg,ibb1,tr3}, Airy~\cite{siv,chen,efr} or Mathieu~\cite{ma,yan} beams. 

Relatively new idea is that of the ``knotted'' light, although topological concepts have long been present in physics~\cite{kelvin,dirac,ab}. The term ``knot'' refers to the characteristics either of electric or magnetic field lines~\cite{ran,ir,besi,kedia,arr,arr2} which can get entangled, or of the nodal lines of the wave intensity or optical vortex lines~\cite{bd,den,bkj,kle,deklerk,su}. It has become possible from the experimental point of view to generate such knotted beams~\cite{leach,leach1,sha,wil}, thus creating the opportunity for practical use.

It is well known that the non-homogeneities of the electric field can provide gradient forces for trapping atoms due to the Stark effect~\cite{dk} or -- equivalently -- due to the polarizability of atoms. This phenomenon provides the basis for a trap called the optical tweezer~\cite{chu,miller}. Assuming the atomic dipole moment to be proportional to the external electric field,
the atomic polarizability $\alpha$ (in general depending on the driving frequency) can be introduced as
\begin{equation}
{\bm d}=\alpha {\bm E}.
\label{de}
\end{equation}
This leads to the equation of motion of an atom in the form
\begin{equation}
m\ddot{\bm r}= ({\bm d}\cdot {\bm \nabla}){\bm E}=\frac{1}{2}\,\alpha {\bm \nabla}(E^2).
\label{rr1}
\end{equation}
From the theory of the Stark effect in atoms it is known that for a blue-detuned beam the polarizability $\alpha$ becomes negative~\cite{odtna}. This causes particles to be dragged into an area with lower value of $E^2$. A natural question arises whether the knotted nodal lines spoken of above can serve as such traps for this kind of particles~\cite{sha}. One might expect that the considered knots constituting regions of weaker field should attract atoms with $\alpha<0$. Similar effect should be observed for charged particles, (e.g., electrons), moving in the ponderomotive potential, where the coefficient $\alpha$ is negative as well~\cite{ibb1}. So, one can ask an important question: can all these particles be forced to move along previously chosen and designed knotted paths? Up to our knowledge, this interesting issue has not been studied so far. Some simpler trajectories as rings or helices have already been shown to be actually realized, for instance in Bessel beams~\cite{tr3}. It seems worthy of some attention to verify whether the knotted structure can be transferred from the field to particles. This might open a variety of applications including the guidance of particles around special kinds of obstacles, the engineering of complex nanoparticles or the generation of knotted nanocircuits (e.g. for electrons that are subject to ponderomotive force).

The manipulation of particles has become an extremely topical and important issue in recent years due to the significant applications in physics, chemistry, biology or medicine (see for instance~\cite{ste,fazal,pad,woe,bowpa,grier1,brad}). The knotted trajectories would constitute a new and potentially widely applicable family and we hope that current work adds a tiny contribution in this regard. Motion of charged particles in the knotted electromagnetic field (but in the sense of the field-lines knots) was dealt with in~\cite{arr1}.

As mentioned above, knots understood as nodal lines create some kind of potential valleys for particles in question. Their guidance along these lines is, however, a highly nontrivial issue, since the tightly knotted lines disturb the structure of these valleys and make them substantially shallower, which can knock particles out of their designed trajectories. In this work it is shown that the particles can still follow knotted paths if the initial conditions are carefully tuned.

Let us express the complex electric field of a monochromatic wave through an envelope $\Psi(\bm{r})$:
\begin{equation}
\bm{E}(\bm{r},t)=\bm{E}_0e^{i(kz-\omega t)}\Psi(\bm{r}),
\label{ep}
\end{equation}
where $\bm{E}_0$ is a constant vector. Below it will be convenient to make use of the dimensionless coordinates
\begin{equation}\label{diml}
\xi_x=k x,\;\;\;\xi_y=k y,\;\;\;\xi=\sqrt{\xi_x^2+\xi_y^2} ,\;\;\;\zeta=k z,
\end{equation}
where $k=\omega/c$. In our approach, the third component plays a special role, so we prefer to denote it with the symbol $\zeta$ instead of $\xi_z$. These  variables turned out to be convenient in our previous papers dealing with trapping of particles by light beams. However, for the convenience of the reader some of the knotted beams finally obtained in the subsequent sections will be given more traditional form as well. 

In what follows the bold mathematical symbols refer to two-dimensional vectors, as for instance $\bm{\xi}=[\xi_x,\xi_y]$.  Similarly $\bm{r}=[x,y]$.

After having averaged the potential on the r.h.s of~(\ref{rr1}) over fast optical oscillations one gets the smoothed equations of motion in the form:
\begin{equation}
\ddot{\bm{\xi}}=-\beta\bm{\partial}_\xi|\Psi|^2,\;\;\;\;\;\; \ddot{\zeta}=-\beta\partial_\zeta|\Psi|^2,
\label{eqmot}
\end{equation}
where $\beta=|\alpha|\cdot |E_0|^2/4mc^2$ and $\partial_x$ denoting $\partial/\partial x$. For laser intensities in the range of $10^7-10^8\,\mathrm{W/cm^2}$ one can achieve the depth of the appropriate potential valley of order of a couple of $\mathrm{meV}$, depending also on the atomic polarizability. The trap then still remains perturbative in the sense that it does not significantly interfere with the internal structure of atomic energy levels. The figures below are of illustrative character and are performed for the value of $\beta$ equal to unity for reasons of clear visualization. For $\beta\ll 1$ the trap is still operative, but the particles require very precise preparation of the initial states particularly with respect to the transverse velocities. For electrons subject to ponderomotive force the value of $\beta$ may be increased which corresponds to proportionally stronger laser fields.

In the following sections we first describe how to theoretically construct Gaussian beams with a given knotted topology and then, using~(\ref{eqmot}), the results of the numerical integration, demonstrating particle trajectories for four special knots (the unknot, the Hopf link, the Borromean rings and the trefoil) are presented.

\section{Gaussian beams with nodal knots}\label{gbk}

One can treat a knot as a closed curve in $\mathbb {R}^3$, i.e., a curve that constitutes a homeomorphic image of $S^1$. It then forms a loop.  This curve may be a nodal line of a certain complex-valued function of the spatial variables $x,y,z$ (in our case $\xi_x,\xi_y,\zeta$). Of course, a knot can also be composed of several disjoint loops, that are tangled up forming a link  (e.g. the Hopf link, the Borromean rings and so on). 

The construction leading to the required specific knots or links can be found elsewhere~\cite{bra,bkj,king,bode}. The details remain beyond the scope of this work and we will limit ourselves to mentioning the main steps only.  First,  one constructs a polynomial $q(u,v)$ of two complex variables $u$ and $v$ which satisfy the condition for the three-dimensional sphere: $|u|^2+|v|^2=1$. The examples of such polynomials are given in the following section.  All points where $q(u,v)=0$ represent an algebraic knot.

Since we are concerned about knots in ${\mathbb R}^3$ rather than on $S^3$, the next step is to use the stereographic projection by means of the relations:
\begin{subequations}\label{stereo}
\begin{align}
u(\bm{\xi},\zeta)&=\frac{\bm{\xi}^2+\zeta^2-1+2 i \zeta}{\bm{\xi}^2+\zeta^2+1},\label{stereou}\\
v(\bm{\xi},\zeta)&=\frac{2(\xi_x+i\xi_y)}{\bm{\xi}^2+\zeta^2+1},\label{stereov}
\end{align}
\end{subequations}
and to require $q(u(\bm{\xi},\zeta),v(\bm{\xi},\zeta))=0$. In that way the knot curve becomes an intersection of two surfaces in three-dimensional space: $\operatorname{Re}q(u(\bm{\xi},\zeta),v(\bm{\xi},\zeta))=0$ and $\operatorname{Im} q(u(\bm{\xi},\zeta),v(\bm{\xi},\zeta))=0$. Since $q(u,v)$ is a polynomial, $q(u(\bm{\xi},\zeta),v(\bm{\xi},\zeta))$ can again be treated as a polynomial (called the Milnor polynomial~\cite{milnor}), upon removing the common denominator stemming from~(\ref{stereo}). For the appropriate Milnor polynomial the symbol $q_M(\bm{\xi},\zeta)$ is reserved below. 

The knot lines obtained that way cannot, however, constitute nodal lines of light waves since, in general, the wave equation would not be satisfied. We are rather interested in special superpositions of Gaussian beams which, on one hand, satisfy the paraxial equation:
\begin{equation}\label{paraxial}
\mathcal{4} \Psi(\bm{\xi},\zeta)+2i\partial_\zeta \Psi({\bm{\xi}},\zeta)=0,
\end{equation}
with $\mathcal{4}$ denoting the two-dimensional Laplace operator in variables $\bm{\xi}$, and on the other, exhibit knotted nodal lines. In order to construct such waves the obvious equivalence of~(\ref{paraxial}) to the two-dimensional Schr\"odinger equation for a free particle
\begin{equation}\label{schr}
-\frac{\hbar^2}{2m}\,\mathcal{4}\Psi(\bm{r},t)=i\hbar\partial_t\Psi(\bm{r},t),
\end{equation}
can be made use of. These two equations become identical upon the identification:
\begin{equation}
mc^2=\hbar\omega,\;\;\;\; \bm{\xi}=\bm{r},\;\;\;\; \zeta=c t.
\label{id}
\end{equation}
Therefore, instead of talking about waves satisfying the paraxial equation one can consider the time evolution of a free-particle's wave-function in two dimensions. The time-dependent (or $\zeta$-dependent) wave-function which coincides with the relevant Milnor polynomial (eventually with a Gaussian factor) at $t=0$ (i.e. on the surface $\zeta=0$) and evolves according to the equation~(\ref{schr}) will inherit from that polynomial the topological structure of the nodal lines.

The evolution of a free particle in quantum mechanics is, obviously, well known and is determined by the Schr\"odinger propagator which, in two spatial dimensions, has the form
\begin{equation}\label{prop}
K({\bm r}, t; {\bm r}',t')=\frac{-im}{2\pi \hbar(t-t')}\,\exp\left[i\,\frac{m({\bm r}-{\bm r}')^2}{2\hbar (t-t')}\right].
\end{equation}
It follows then that the function $\Psi(\bm{\xi},\zeta)$
\begin{equation}\label{evol}
\Psi(\bm{\xi},\zeta)=\int d^2\xi' K({\bm \xi}, \zeta; {\bm \xi}',0) e^{-\kappa \bm{\xi}'^2}q_M(\bm{\xi}',0),
\end{equation}
where $\kappa>0$ and
\begin{equation}\label{prop1}
K({\bm \xi}, \zeta; {\bm \xi}',\zeta')=\frac{-i}{2\pi (\zeta-\zeta')}\,\,\exp\left[i\,\frac{({\bm \xi}-{\bm \xi}')^2}{2(\zeta-\zeta')}\right],
\end{equation}
fulfills the paraxial equation~(\ref{paraxial}), while maintaining the desired knotted structure of the nodal lines and the Gaussian character (at least for a mild Gaussian). Apparently $K({\bm \xi}, \zeta; {\bm \xi}',\zeta')$ does not satisfy the paraxial condition due to the divergence as $\zeta\rightarrow \zeta'$, but one should remember that this behavior is smoothed out thanks to the integration in~(\ref{evol}).

\section{Knotted trajectories of particles}
\label{kta}

In this section four specific examples of such knotted lines are dealt with: the unknot (i.e. the ring), the Hopf link, the Borromean rings and the trefoil.
The first three examples are generated from he polynomial $q(u,v)$, which can be written in the general form
\begin{equation}
q(u,v)=\prod\limits_{k=0}^{n-1} (u-\varepsilon_n^{(k)}v),
\label{qgen}
\end{equation}
where $\varepsilon_n^{(k)}$, $k=0,1,2,\ldots,n-1$, denote the subsequent $n$th roots of unity.

\subsection{The unknot}\label{unknot}
The polynomial $q(u,v)$ to generate a ring is obtained by inserting $n=1$ into~(\ref{qgen}), which leads to
\begin{equation}
q(u,v)=u-v.
\label{qring}
\end{equation}
Substituting $u$ and $v$ in the form of~(\ref{stereo}), the Milnor polynomial at $\zeta=0$ is obtained as
\begin{equation}
q_M(\bm{\xi},0)=-1+\xi_x^2+\xi_y^2-2(\xi_x+i\xi_y).
\label{fring}
\end{equation}
In order to keep the possibility of modifying the spatial size of the knot, here, and in the following examples, an additional scaling factor $\gamma$ will be introduced wherever the powers $\xi_x$ and $\xi_y$ occur. This procedure does not change the topology of the knot, but offers the possibility to adjust its size to the dimensions achievable in an experiment. Therefore, instead of~(\ref{fring}) we will use
\begin{equation}
q_M(\bm{\xi},0)=-1+\gamma^2(\xi_x^2+\xi_y^2)-2\gamma(\xi_x+i\xi_y).
\label{fringg}
\end{equation}
This leads to the paraxial envelope in the form
\begin{eqnarray}
\Psi(\bm{\xi},\zeta)&\!\! =&\!\! \frac{-i}{2\pi\zeta}\int\limits_0^\infty d\xi' \xi'\int\limits_0^{2\pi} d\phi' e^{\frac{i}{2\zeta}(\xi^2+\xi'^2-2\xi\xi'\cos(\phi-\phi'))}\nonumber\\
&&\!\!\times e^{-\kappa \xi'^2}(-1+\gamma^2\xi'^2-2\gamma\xi' e^{i\phi'}),\label{frev}
\end{eqnarray}
where polar coordinates have been introduced. The subsequent integrals over $\phi'$ and $\xi'$ can be easily calculated with the use of the formulas of Appendix, with the following result
\begin{eqnarray}
\Psi(\bm{\xi},\zeta)&\!\! =&\!\! \frac{i}{\zeta}\,e^{\frac{i}{2\zeta}\,\xi^2}\int\limits_0^\infty d\xi' e^{-(\kappa-\frac{i}{2\zeta})\xi'^2}\label{frev1}\\
&&\!\!\!\!\!\times \left[\xi'(1-\gamma^2\xi'^2)J_0(\xi\xi'/\zeta)-2ie^{i\phi}\gamma\xi'^2J_1(\xi\xi'/\zeta)\right],\nonumber
\end{eqnarray}
with $J_n$ denoting the Bessel functions, and then
\begin{equation}
\Psi(\bm{\xi},\zeta)=e^{-\frac{\kappa\xi^2}{c(\zeta)}}\,\left(\frac{-1}{c(\zeta)}+\frac{2 i \gamma^2\zeta}{c(\zeta)^2}+\frac{\gamma^2\xi^2}{c(\zeta)^3}-\frac{2\gamma\xi}{c(\zeta)^2}\,e^{i\phi}\right),\label{frev3}
\end{equation}
where $c(\zeta)=2 i \kappa\zeta +1$. As can be easily verified, the function $\Psi(\bm{\xi},\zeta)$ obtained above satisfies the paraxial equation~(\ref{paraxial}). The role of the factors $\gamma$ introduced in~(\ref{fringg}) merely reduces to fixing the relative intensities when superimposing various Gaussian beams in~(\ref{frev3}). This expression may be rewritten in more traditional but less convenient indications, if it is noticed that
\begin{subequations}\label{nota}
\begin{align}
c(\zeta)&=1+i\frac{z}{z_R},\label{nota1}\\
\kappa&=\frac{1}{k^2w_0^2},\label{nota2}\\
\frac{1}{c(\zeta)}&=\frac{w_0}{w(z)}e^{-i\psi(z)},
\end{align}
\end{subequations}
where $z_R$ is the Rayleigh length, $w_0$ denotes the beam waist, $w(z)=w_0\sqrt{1+(z/z_R)^2}$ is the beam radius, $R(z)=z(1+(z_R/z)^2)$ stands for the wavefront curvature and $\psi(z)=\arctan (z/z_R)$ is the  Gouy phase.
Expression ~(\ref{frev3}) then takes the form ($\bm{r}$ refers to the transverse coordinates only):
\begin{eqnarray}
\Psi(\bm{r},z)&&\!\!\!\!\!\!=\frac{w_0}{w(z)}\exp\left[-\frac{r^2}{w(z)^2}-i\frac{r^2 k}{2 R(z)}-i\psi(z)\right]\label{frev3s}\\
&&\!\!\!\!\!\!\!\!\!\!\!\!\times\left(-1+\frac{2 i\gamma kz}{1+iz/z_R}+\frac{\gamma^2 k^2 r^2}{(1+iz/z_R)^2}-\frac{2\gamma kre^{i\phi}}{1+iz/z_R}\right),\nonumber
\end{eqnarray}
This beam belongs to the wide class of the so called Hypergeometric-Gaussian beams, theoretically described in~\cite{karimi} and experimentally created with the use of computer-generated hologram and spatial light modulator. All beams representing more complicated knots obtained in the following subsections can be given the similar form.

Coming back to~(\ref{frev3}), one should have a look onto two special limits. 
For $\kappa\rightarrow 0$ one has $c(\zeta)\rightarrow 1$, and the appropriate paraxial polynomial is recovered
\begin{equation}
q_p(\bm{\xi},\zeta)=-1+2i\gamma^2\zeta+\gamma^2\xi^2-2\gamma(\xi_x+i\xi_y).
\label{parrin}
\end{equation}
It still constitutes the solution of the paraxial equation and possesses the same knotted structure, but cannot represent the true light wave due to its spatial divergence at infinity. In turn, letting $\zeta\rightarrow 0$ the function (\ref{fringg}) tempered with the Gaussian is obtained. This is obvious from the very construction of the envelope $\Psi$, since the Schr\"odinger propagator satisfies the condition
\begin{equation}
\lim\limits_{t\rightarrow t'}K({\bm r}, t; {\bm r}',t')=\delta^{(2)}({\bm r}-{\bm r'}).
\label{lk}
\end{equation}

\begin{figure}[h]
\begin{center}
\includegraphics[width=0.5\textwidth,angle=0]{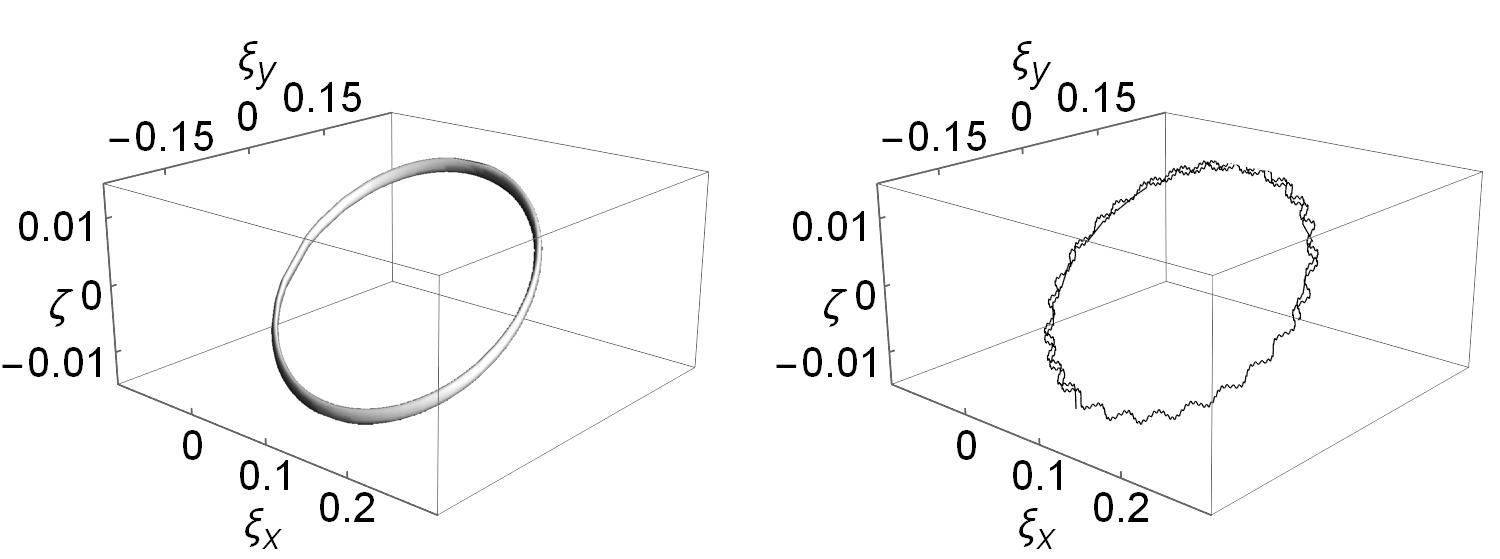}
\end{center}
\caption{The nodal line of the light wave representing the simplest ``knot'' (left plot), and the trajectory of the injected particle (right plot). The values of parameters are: $\kappa = 0.01, \beta=1, \gamma=10$. According to~(\ref{diml}) the units on axes correspond to $\lambda/2\pi$.}
\label{plring}
\end{figure}

With this form of $\Psi(\bm{\xi},\zeta)$ the equation of motion~(\ref{eqmot}) of the particle can be numerically solved. The results are presented in Fig.~\ref{plring}. It is visible that the particle exactly follows the nodal line of the wave, thereby moving along the knotted trajectory. This trajectory is limited to the size of order $0.02\,\lambda$, which justifies the adoption of the paraxial approximation. If necessary, this area can be further reduced by increasing the value of $\gamma$, but the drawings for more complicated knots would then become less transparent.  The chosen exemplary value of $\kappa=0.01$ corresponds to $w_0\approx 1.6\lambda$, which stays in agreement with the approximations used~\cite{st}. Moreover, it may be changed within wide limits, without significant modifications of the outcome.

Apart from the motion along the nodal line of the electromagnetic wave the particle performs an oscillatory motion perpendicular to it. The amplitude of these oscillations depends on the depth of the binding-potential valley, i.e. on the intensity of the wave, and on the initial tuning of the position and velocity.

\subsection{The Hopf link}\label{hopf}

In order to obtain the Hopf link, one has to set $n=2$ in~(\ref{qgen}), obtaining
\begin{equation}
q(u,v)=(u-v)(u+v).
\label{qhopf}
\end{equation}
The appropriate Milnor polynomial, found by substituting $u$ and $v$ according to~(\ref{stereo}) and reduced to the plane $\zeta=0$, has now the form:
\begin{equation}
q_M(\bm{\xi},0)=(1-\xi_x^2-\xi_y^2)^2-4(\xi_x+i\xi_y)^2.
\label{fhopf}
\end{equation}
Following the procedure outlined in the case of the unknot and introducing the scale parameter $\gamma$, one finds
\begin{eqnarray}
\Psi(\bm{\xi},\zeta)&\!\! =&\!\! \frac{-i}{2\pi\zeta}\int\limits_0^\infty d\xi' \xi'\int\limits_0^{2\pi} d\phi' e^{\frac{i}{2\zeta}(\xi^2+\xi'^2-2\xi\xi'\cos(\phi-\phi'))}\nonumber\\
&&\!\!\times e^{-\kappa \bm{\xi}'^2}\big[(1-\gamma^2\xi'^2)^2-4\gamma^2\xi'^2 e^{2i\phi'}\big],\label{fhev}
\end{eqnarray}
and consequently
\begin{eqnarray}
\Psi(\bm{\xi},\zeta)&\!\! =&\!\! \frac{-i}{\zeta}\,e^{\frac{i}{2\zeta}\,\xi^2}\int\limits_0^\infty d\xi' e^{-(\kappa-\frac{i}{2\zeta})\xi'^2}\big[\xi'(1-\gamma^2\xi'^2)^2)\nonumber\\
&&\!\!\times J_0(\xi\xi'/\zeta)+4e^{2i\phi}\gamma^2\xi'^2J_2(\xi\xi'/\zeta)\big],\label{fhev1}
\end{eqnarray}
where again the integrals collected in the Appendix have been used. This leads to the paraxial wave-function
\begin{eqnarray}
\Psi(\bm{\xi},\zeta)&\!\! =&\!\!e^{-\frac{\kappa\xi^2}{c(\zeta)}}\,\bigg(\frac{1}{c(\zeta)}-\frac{4 i\gamma^2 \zeta}{c(\zeta)^2}-\frac{2\gamma^2\xi^2}{c(\zeta)^3}-\frac{8\gamma^4\zeta^2}{c(\zeta)^3}\nonumber\\
&&\!\!+\frac{8i\gamma^4\zeta\xi^2}{c(\zeta)^4}+\frac{\gamma^4\xi^4}{c(\zeta)^5}-\frac{4\gamma^2\xi^2}{c(\zeta)^3}\,e^{2i\phi}\bigg),\label{fhev3}
\end{eqnarray}
possessing nodal lines representing the Hopf link. Using the relations~(\ref{nota}) the expression~(\ref{fhev3}) can be easily given the form of a superposition of the Hypergeometric-Gaussian modes similar to~(\ref{frev3s}):
\begin{eqnarray}
\Psi(\bm{r},z)&&\!\!\!\!\!\!=\frac{w_0}{w(z)}\exp\left[-\frac{r^2}{w(z)^2}-i\frac{r^2 k}{2 R(z)}-i\psi(z)\right]\label{fhev3s}\\
&&\!\!\!\!\!\!\!\!\times\bigg(1-\frac{4 i\gamma kz}{1+iz/z_R}-\frac{2\gamma^2 k^2 r^2}{(1+iz/z_R)^2}-\frac{8\gamma^4 k^2 z^2}{(1+iz/z_R)^2}\nonumber\\
&&\!\!\!\!\!\!\!\!+\frac{8 i\gamma^4 k^3r^2z}{(1+iz/z_R)^3}+\frac{\gamma^4 k^4 r^4}{(1+iz/z_R)^4}-\frac{4\gamma^2 k^2r^2e^{2i\phi}}{(1+iz/z_R)^2}\bigg).\nonumber
\end{eqnarray}

If $\kappa\rightarrow 0$, the paraxial polynomial is obtained as
\begin{eqnarray}
q_p(\bm{\xi},\zeta)&\!\! =&\!\!1-4i\gamma^2\zeta-2\gamma^2(\xi_x^2+\xi_y^2)-8\gamma^4\zeta^2\label{parhop}\\
&&\!\!\!\!\!\!\!\!\!+8i\gamma^4\zeta (\xi_x^2+\xi_y^2)+\gamma^4(\xi_x^2+\xi_y^2)^2-4\gamma^2(\xi_x+i\xi_y)^2. \nonumber
\end{eqnarray}

\begin{figure}[h]
\begin{center}
\includegraphics[width=0.5\textwidth,angle=0]{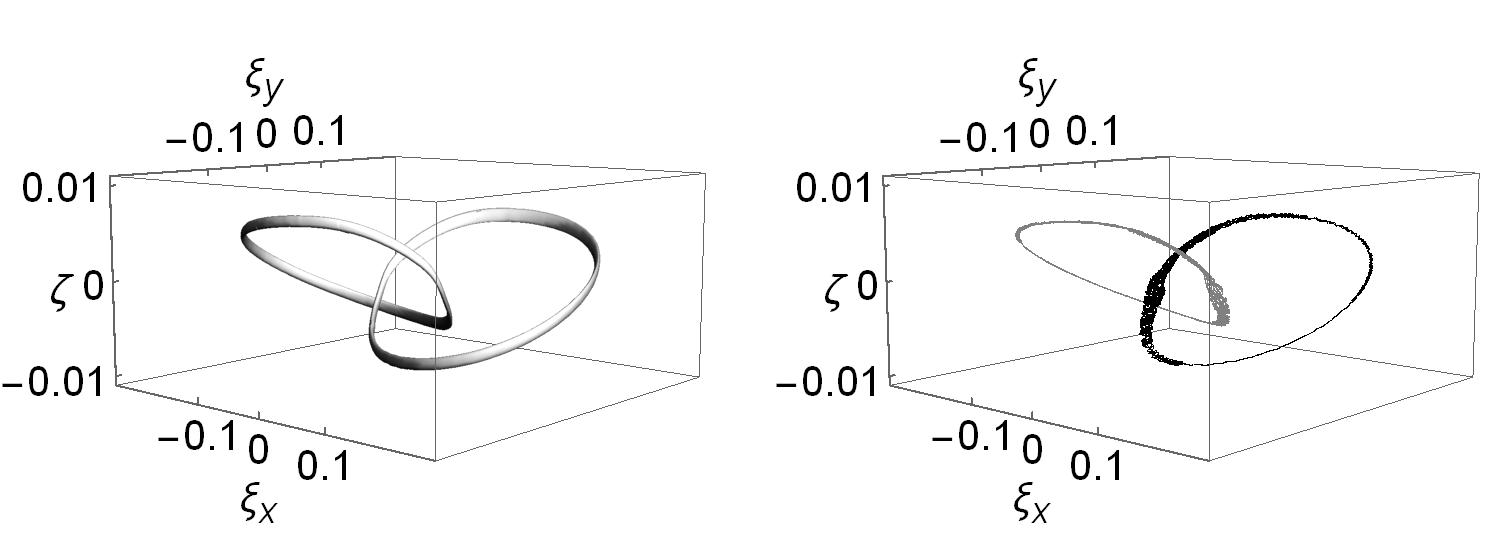}
\end{center}
\caption{The nodal line of the light wave representing the Hopf link (left plot), and the trajectories of two injected particles (right plot). The values of parameters are the same as in Fig.~\ref{plring}.}
\label{plhopf}
\end{figure}

The trajectories of two particles moving according to Eqs~(\ref{eqmot}) with $\Psi(\bm{\xi},\zeta)$ given by~(\ref{fhev3}) are drawn in Fig.~\ref{plhopf}. As can be seen, each particle follows one of the two rings constituting the Hopfian.

Looking closely at the trajectories one can again recognize the oscillatory motion in perpendicular directions. Upon precise examination, the amplitude of these oscillations turns out to increase in places where both rings are passing each other (the apparent broadening of the trajectory appears): this is due to the local flattening of the particle-binding potential. These are also places where a possible jump of the particle between the rings can eventually occur if the initial conditions are not sufficiently tuned. For more complex knots, this effect sets higher requirements regarding the preparation of the initial states of the particles. For instance in the case of the trefoil knot, spoken of in Sec.~\ref{trefoil}, the initial perpendicular velocity could reach $2\times 10^{-4}\, \mathrm{c}$ provided the value of the parameter $\beta$ is set to unity. Lowering this value for instance by $2n$ orders of magnitude requires reducing the above velocity by $n$ orders. For laser intensities spoken of in the introduction, the energies of the perpendicular motion should be reduced to hundreds of nanoelectronvolts.

\subsection{The Borromean rings}\label{borings}

The link known as the Borromean rings is composed of three loops, and, therefore, one has to set $n=3$ in~(\ref{qgen}), obtaining
\begin{equation}
q(u,v)=(u-v)(u-e^{2\pi i/3}v)(u-e^{4\pi i/3}v).
\label{qbor}
\end{equation}
Consequently
\begin{equation}
q_M(\bm{\xi},0)=(-1+\xi_x^2+\xi_y^2)^3-8(\xi_x+i\xi_y)^3,
\label{fbor}
\end{equation}
and, similarly as before,
\begin{eqnarray}
\Psi(\bm{\xi},\zeta)&\!\! =&\!\! \frac{-i}{2\pi\zeta}\int\limits_0^\infty d\xi' \xi'\int\limits_0^{2\pi} d\phi' e^{\frac{i}{2\zeta}(\xi^2+\xi'^2-2\xi\xi'\cos(\phi-\phi'))}\nonumber\\
&&\!\!\times e^{-\kappa \bm{\xi}'^2}[(-1+\gamma^2\xi'^2)^3-8\gamma^3\xi'^3 e^{3i\phi'}].\label{fbev}
\end{eqnarray}
Using integrals listed in Appendix, we first come to
\begin{eqnarray}
\Psi(\bm{\xi},\zeta)& \!\!=&\!\! \frac{-i}{\zeta}\,e^{\frac{i}{2\zeta}\,\xi^2}\int\limits_0^\infty d\xi' e^{-(\kappa-\frac{i}{2\zeta})\xi'^2}\big[\xi'(-1+\gamma^2\xi'^2)^3\nonumber\\
&&\!\!\times J_0(\xi\xi'/\zeta)-8ie^{3i\phi}\gamma^3\xi'^3J_3(\xi\xi'/\zeta)\big].\label{fbev1}
\end{eqnarray}
and finally get the paraxial envelope as
\begin{eqnarray}
\Psi(\bm{\xi},\zeta)&\!\! =&\!\!e^{-\frac{\kappa\xi^2}{c(\zeta)}}\,\bigg(\frac{-1}{c(\zeta)}+\frac{6 i \gamma^2\zeta}{c(\zeta)^2}+\frac{3\gamma^2\xi^2}{c(\zeta)^3}+\frac{24\gamma^4\zeta^2}{c(\zeta)^3}\nonumber\\
&&\!\!-\frac{48i\gamma^6\zeta^3}{c(\zeta)^4}-\frac{24i\gamma^4\xi^2\zeta}{c(\zeta)^4}-\frac{72\gamma^6\xi^2\zeta^2}{c(\zeta)^5}\label{fbev3}\\
&&\!\!-\frac{3\gamma^4\xi^4}{c(\zeta)^5}+\frac{18i\gamma^6\xi^4\zeta}{c(\zeta)^6}+\frac{\gamma^6\xi^6}{c(\zeta)^7}-\frac{8\gamma^3\xi^3}{c(\zeta)^4}\,e^{3i\phi}\bigg).\nonumber
\end{eqnarray}
This envelope could again be represented as a combination of Hypergeometric-Gaussian beams in an obvious way but there is no need to write down the explicit formula here (and after formula~(\ref{fbev4})).

The corresponding paraxial polynomial has the form
\begin{eqnarray}
q_p(\bm{\xi},\zeta)& \!\!=&\!\!-1+6i\gamma^2\zeta+3\gamma^2(\xi_x^2+\xi_y^2)+24\gamma^4\zeta^2 \label{parbor}\\
&&\!\!-48i\gamma^6\zeta^3-24i\gamma^4 (\xi_x^2+\xi_y^2)\zeta-72 \gamma^6(\xi_x^2+\xi_y^2)\zeta^2\nonumber\\
&&\!\!-3\gamma^4(\xi_x^2+\xi_y^2)^2+18i\gamma^6(\xi_x^2+\xi_y^2)^2\zeta\nonumber\\
&&\!\!+\gamma^6(\xi_x^2+\xi_y^2)^3-8\gamma^3(\xi_x+i\xi_y)^3.\nonumber
\end{eqnarray}

\begin{figure}[h]
\begin{center}
\includegraphics[width=0.5\textwidth,angle=0]{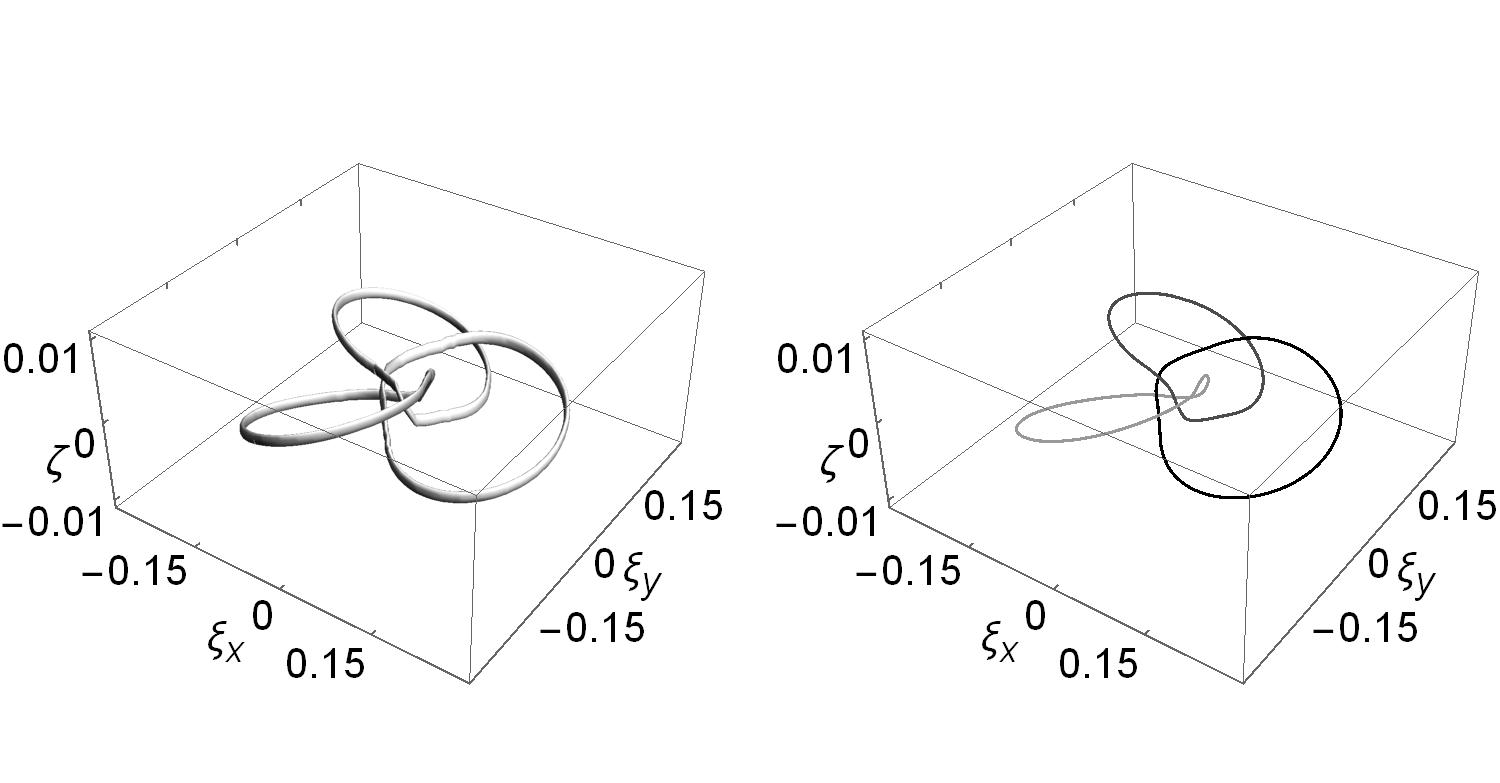}
\end{center}
\caption{The nodal line of the light wave representing the Borromean rings (left plot), and the trajectories of three injected particles (right plot). The values of parameters are the same as before.}
\label{plbor}
\end{figure}

Now one can pass to the motion of atoms accurately injected into the electromagnetic field~(\ref{fbev3}). The numerical calculations show again that trajectories of three particles, presented in Fig.~\ref{plbor}, replicate the knotted structure, although it is much more challenging from the numerical (and consequently -- experimental) point of view, due to the presence of almost ``intersecting'' lines.

The broadening of the trajectories of particles close to the passing points is less visible and depends on the precision of tuning their initial states, which has been successfully implemented in the numerical calculations. Due to the complicated nodal structure, the potential becomes relatively shallow (and with many local minima). The initial conditions satisfied by particles have to be fine tuned so as to appropriately place them in the field and to avoid chaotic motion. This is reflected in the numerical calculations, where trajectories become sensitive to the initial states.

\subsection{The trefoil}\label{trefoil}

A very nontrivial knot known as the trefoil is generated from a polynomial other than the types described with the formula~(\ref{qgen}). This time it has the form
\begin{equation}
q(u,v)=u^2-v^3.
\label{qtre}
\end{equation}
leading to
\begin{equation}
q_M(\bm{\xi},0)=(1+\xi_x^2+\xi_y^2)(1-\xi_x^2-\xi_y^2)^2-8(\xi_x+i\xi_y)^3,
\label{ftre}
\end{equation}
and belongs to the larger family of knots obtained from the expression
\begin{equation}
q(u,v)=u^2-v^n, 
\label{qtren}
\end{equation}
with $n\in \mathbb{N}$~\cite{king}, like the cinqefoil knot ($n=5$) or the septafoil knot ($n=7$) etc.

The wave envelope is constructed from $q_M(\bm{\xi},0)$ as in the previous subsections. We define 
\begin{eqnarray}
\Psi(\bm{\xi},\zeta)&&\!\!\!\!\!\! = \frac{-i}{2\pi\zeta}\int\limits_0^\infty d\xi' \xi'\int\limits_0^{2\pi} d\phi' e^{\frac{i}{2\zeta}(\xi^2+\xi'^2-2\xi\xi'\cos(\phi-\phi'))}\nonumber\\
&&\!\!\!\!\!\!\!\!\!\!\times e^{-\kappa \bm{\xi}'^2}[(1+\gamma^2\xi'^2)(1-\gamma^2\xi'^2)^2-8\gamma^3\xi'^3 e^{3i\phi'}].\label{ftrep}
\end{eqnarray}
and integrate first with respect to $\phi'$:
\begin{eqnarray}
\Psi(\bm{\xi},\zeta) = &&\!\!\!\!\!\!\frac{-i}{\zeta}\,e^{\frac{i}{2\zeta}\,\xi^2}\int\limits_0^\infty d\xi' e^{-(\kappa-\frac{i}{2\zeta})\xi'^2}\big[\xi'(1+\gamma^2\xi'^2)\label{fbev2}\\
&&\!\!\!\!\!\!\!\!\!\!\!\!\times (1-\gamma^2\xi'^2)^2J_0(\xi\xi'/\zeta)-8ie^{3i\phi}\gamma^3\xi'^3J_3(\xi\xi'/\zeta)\big].\nonumber
\end{eqnarray}
and finally over $\xi'$:
\begin{eqnarray}
\Psi(\bm{\xi},\zeta)&\!\! =&\!\!e^{-\frac{\kappa\xi^2}{c(\zeta)}}\,\bigg(\frac{1}{c(\zeta)}-\frac{2 i \gamma^2\zeta}{c(\zeta)^2}-\frac{\gamma^2\xi^2}{c(\zeta)^3}+\frac{8\gamma^4\zeta^2}{c(\zeta)^3}\nonumber\\
&&\!\!-\frac{8i\gamma^4\xi^2\zeta}{c(\zeta)^4}-\frac{48i\gamma^6\zeta^3}{c(\zeta)^4}-\frac{\gamma^4\xi^4}{c(\zeta)^5}-\frac{72\gamma^6\xi^2\zeta^2}{c(\zeta)^5}\nonumber\\
&&+\frac{18i\gamma^6\xi^4\zeta}{c(\zeta)^6}+\frac{\gamma^6\xi^6}{c(\zeta)^7}-\frac{8\gamma^3\xi^3}{c(\zeta)^4}\,e^{3i\phi}\bigg).\label{fbev4}
\end{eqnarray}

The corresponding paraxial polynomial may be obtained in the form
\begin{eqnarray}
q_p(\bm{\xi},\zeta)&\!\! =&\!\!1-2i\gamma^2\zeta-\gamma^2(\xi_x^2+\xi_y^2)+8\gamma^4\zeta^2\nonumber \\
&&\!\!-8i \gamma^4(\xi_x^2+\xi_y^2)\zeta-48i\gamma^6\zeta^3-\gamma^4(\xi_x^2+\xi_y^2)^2\nonumber\\
&&\!\!-72 \gamma^6(\xi_x^2+\xi_y^2)\zeta^2+18i\gamma^6(\xi_x^2+\xi_y^2)^2\zeta\nonumber\\
&&\!\!+\gamma^6(\xi_x^2+\xi_y^2)^3-8\gamma^3(\xi_x+i\xi_y)^3.\label{partre}
\end{eqnarray}

\begin{figure}[h]
\begin{center}
\includegraphics[width=0.5\textwidth,angle=0]{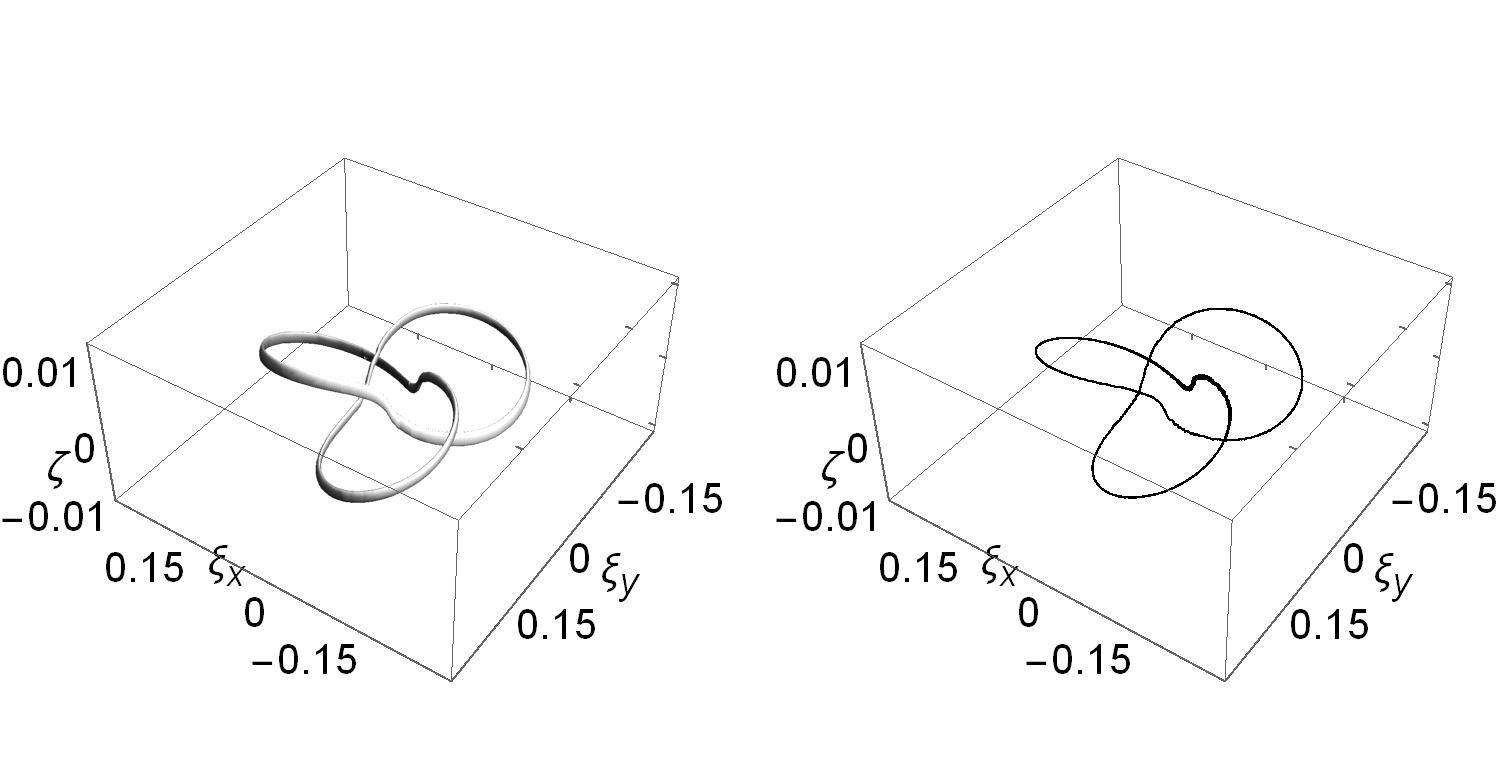}
\end{center}
\caption{The nodal line of the light wave representing the trefoil (left plot), and the trajectory of the injected particle (right plot). The values of parameters are the same as before.}
\label{pltre}
\end{figure}

The results of the numerical calculations of the atom trajectory in this case are presented in Fig.~\ref{pltre}. It exactly follows the trefoil line, again performing the perpendicular oscillatory motion. The conclusions are similar to those of the Borromean rings. The comparable effects can be obtained for the cinquefoil and other knots.

\section{Summary}\label{sum}

In conclusion, the paper presents a simple analytical way of obtaining Gaussian light beams with a given knotted  topology of the nodal lines starting from the Milnor polynomials with an additional scaling factor. The method is based on the similarity between the paraxial wave equation and the two-dimensional Schr\"odinger equation. Then it is shown in a numerical way that neutral, polarizable particles, such as atoms, but also charged particles (electrons) subject to ponderomotive potential can be forced to follow these knot lines. In all considered cases this result turned out to be feasible upon very precise tuning of the initial positions and velocities of particles. The more braided nodal structure, the shallower potential is created (especially close to the passing points), more redundant minima appear and more accurate preparation of particles is required. The results of this work seem applicable for instance for precise guiding of particles along prescribed complicated pathways, particularly with the presence of obstacles, in the generation of knotted nanocircuits or engineering complex nanoparticles.

\appendix*
\section{Appendix}\label{app}

In this appendix several integrals needed to find knotted solutions of the paraxial equation are collected.

\begin{eqnarray}
\int\limits_0^{2\pi}d\phi' e^{-i\beta\cos(\phi-\phi')} &&e^{in\phi'}= 2\pi (-i)^ne^{i n\phi}J_n(\beta), \nonumber\\
&&\mathrm{for}\;\;\; n=0,1,2,\ldots
\label{ifp}
\end{eqnarray}
\begin{equation}
\int\limits_0^\infty dx\, x\, e^{-a x^2}J_0(bx)=\frac{1}{2a}\,e^{-\frac{b^2}{4a}},\label{bes1}
\end{equation}
\begin{equation}
\int\limits_0^\infty dx\, x^3\, e^{-a x^2}J_0(bx)=\left(\frac{1}{2a}-\frac{b^2}{8a^3}\right)\,e^{-\frac{b^2}{4a}},\label{bes2}
\end{equation}
\begin{equation}
\int\limits_0^\infty dx\, x^5\, e^{-a x^2}J_0(bx)=\left(\frac{1}{a^3}-\frac{b^2}{2a^4}+\frac{b^4}{32a^5}\right)\,e^{-\frac{b^2}{4a}},\label{bes3}
\end{equation}
\begin{equation}
\int\limits_0^\infty dx\, x^2\, e^{-a x^2}J_1(bx)=\frac{b}{4a^2}\,e^{-\frac{b^2}{4a}},\label{bes4}
\end{equation}
\begin{equation}
\int\limits_0^\infty dx\, x^3\, e^{-a x^2}J_2(bx)=\frac{b^2}{8a^3}\,e^{-\frac{b^2}{4a}},\label{bes5}
\end{equation}
\begin{equation}
\int\limits_0^\infty dx\, x^4\, e^{-a x^2}J_3(bx)=\frac{b^3}{16a^4}\,e^{-\frac{b^2}{4a}}.\label{bes6}
\end{equation}

\end{document}